\documentclass[prl,aps,twocolumn,groupedaddress,superscriptaddress,floatfix,showpacs]{revtex4}
\usepackage{epsfig}
\usepackage{mathrsfs}
\usepackage{amsmath}
\usepackage{textcmds}
\usepackage{amssymb}
\usepackage{mathptmx}
\allowdisplaybreaks[1]
\newcommand{\be}{\begin{equation}}
\newcommand{\ee}{\end{equation}}
\newcommand{\bea}{\begin{eqnarray}}
\newcommand{\eea}{\end{eqnarray}}
\newcommand{\ket}[1]{\left|#1\right\rangle}
\newcommand{\bra}[1]{\left\langle #1\right|}

\newcommand{\bc}{\begin{center}}
\newcommand{\ec}{\end{center}}

\renewcommand{\(}{\left(}
\renewcommand{\)}{\right)}
\renewcommand{\[}{\left[}
\renewcommand{\]}{\right]}
\newcommand{\forget}[1]{}

\newcommand{\re}{{\rm e}}

\newcommand{\ri}{{\rm i\,}}
\begin{document}
\title{A Bootstrapping Approach for Generating Maximally Path-Entangled Photon States}
\author{Kishore T. Kapale}
\email{KT-Kapale@wiu.edu}
\affiliation{Department of Physics, Western Illinois University, Macomb, IL 61455-1367}
\affiliation{Hearne Institute for Theoretical Physics, Department of Physics \& Astronomy,
Louisiana State University,
Baton Rouge, Louisiana 70803-4001}
\author{Jonathan P. Dowling}
\affiliation{Hearne Institute for Theoretical Physics, Department of Physics \& Astronomy,
Louisiana State University,
Baton Rouge, Louisiana 70803-4001}
\begin{abstract}
We propose a {\em bootstrapping} approach to generation of maximally path-entangled states of photons, so called ``NOON states''. Strong atom-light interaction of cavity QED can be employed to generate NOON states with about 100 photons; which can then be used to boost the existing experimental Kerr nonlinearities based on quantum coherence effects to facilitate NOON generation with arbitrarily large number of photons all within the current experimental state of the art technology.  We also offer an alternative scheme that uses an atom-cavity dispersive interaction to obtain sufficiently high Kerr-nonlinearity necessary for arbitrary NOON generation.
\end{abstract}
\maketitle
NOON states, path-entangled states of N photons of the form $\ket{N0::0N}\equiv(\ket{N}_{\rm a} \ket{0}_{\rm b} + \ket{0}_{\rm a} \ket{N}_{\rm b})/\sqrt{2}$, are important for quantum lithography~\cite{QL} and Heisenberg limited interferometry with photons~\cite{HLI}. Several theoretical proposals exist for NOON-state generation; nevertheless, experimentally it seems to be a formidable task. So far, NOON states with only three and four photons have been generated~\cite{NOONExp}. There exists a proposal that simulates a six-photon NOON state by post-selection~\cite{Resch:2005} and hence may not be directly useful for the quantum lithography application. In this context it is imperative to develop practical strategies for generation of high-NOON states. In this letter, we propose a couple of routes to increase the number of photons entangled in the NOON form.

To recollect, existing proposals for generation of maximally path entangled states of photons use either the Linear Optical Quantum Computing (LOQC) approach or some kind of optical nonlinearity. We note that LOQC approaches would be unsuitable in this quest as the larger the required number of particles in the entangled state the lower is the success probability~\cite{Kok:2002}. Thus, the routes using optical nonlinearities seem to be promising in the longer run. Nevertheless, experimental nonlinearities are not as strong as one would require them for the present task, even in the case where quantum coherence effects such as Eletromagnetically Induced Transparency~\cite{EITPhysicsToday}, are employed as discussed later. Here we propose a {\em bootstrapping} approach such that existing experimental nonlinearities could be boosted in order to eventually have a large number of particles in the generated NOON state. 

\begin{figure}[ht]
\centerline{\includegraphics[width=0.96\columnwidth]{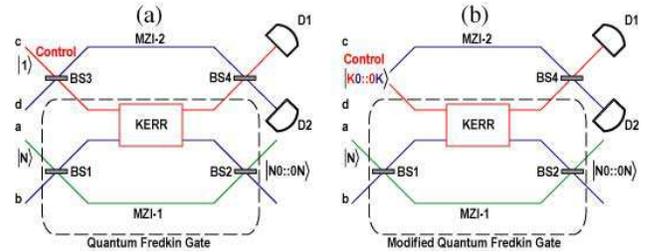}}
\caption{\label{Fig:QFG}Nonlinear process  to generate a high NOON state $\ket{N0::0N}_{a,b}$ via a cross-Kerr nonlinearity. (a) The scheme of Ref. \cite{Gerry:2001} using a single photon Fock state $\ket{1}$ as a control. (b) The Bootstrapping procedure employing a small NOON state $\ket{K0::0K}_{c,d}$ as a control.}
\end{figure}

The bootstrapping technique proposed here involves preparation of a NOON state with a small number of photons (up to 100), which can be used to  boost the conditional nonlinearities necessary to obtain NOON states with an arbitrarily large number of photons $N$ provided, the $N$-photon Fock states are available. The approach is primarily based on the scheme described in Fig.~\ref{Fig:QFG}(a) proposed by Gerry and Campos~\cite{Gerry:2001}. The scheme involves two Mach-Zehnder interferometers (MZI) coupled via a cross-Kerr nonlinearity that can be obtained via schemes based on quantum coherence effects~\cite{PhaseoniumCrossKerr}. Presence of a single photon in mode c is required to give phase shift of $\pi$ to photons in mode b; this phase shift is however not within the reach of current experimental cross-Kerr nonlinearities, to obtain NOON states in the modes a and b at the output. The best current experimental cross-Kerr phase shifts are of the order of 0.1 radians, obtained via atom-light interaction in systems using quantum coherence effects~\cite{CrossKerrExp}. This suggests that further enhancement in the nonlinearity---by roughly  two orders of magnitude---is necessary for NOON state generation. It is, however, important to note  that the scheme requires a conditional phase shift of either 0 or $\pi$, which would require a NOON state (of about $K\approx10\pi$ photons) entering at the input of MZI-2 to act as a control. The necessary setup is shown in Fig.~\ref{Fig:QFG}(b). The presence of a large number of photons would boost the quantum-coherence-based cross-Kerr nonlinearity; enough to obtain a phase shift of $0$ or $\pi$ if all the $K$ control photons occupy mode d or c, all at once. Once the enhanced nonlinearity is used, measurement of the number of photons $n_{\rm D1}$ and $n_{\rm D2}$ would give the output state
$[(-\ri)^{n_{\rm D2}} \ket{N_{\rm a} 0_{\rm b}} + \ri (-\ri)^{n_{\rm D1}} \ket{0_{\rm a} N_{\rm b}}]\sqrt{2}$ which can be trivially corrected for the relative phase of the two components to obtain a NOON state. The condition $n_{\rm D1} + n_{\rm D2}=K$ if satisfied means a successful NOON generation even with inefficient detectors provided they are photon number resolving~\cite{PhotonNumberResolvingDetectors}.

\begin{figure}[ht]
\vskip0.5cm
\includegraphics[width=0.96\columnwidth]{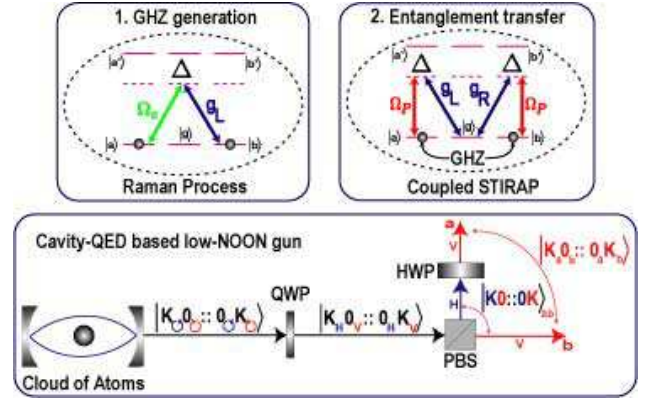}
\caption{\label{Fig:NOONGun} A NOON Gun for small number of photons ($K \approx 100$): A scheme to transfer GHZ state of atoms into a maximally entangled state of photons into two circular polarization modes via collective interaction of atoms with light fields. By using usual optical elements such as Quarter-Wave Plate (QWP), Polarizing Beam Splitter (PBS) and Half-Wave Plate (HWP) the polarization entangled photons in the two circular polarization modes emitted by the cavity  can be converted to path entangled NOON state. The insets show the two important steps 1.~preparation of atoms in the GHZ state, and 2.~transfer of entanglement from atoms to the entangled polarization modes of the cavity.}
\end{figure}

As pointed out earlier, the foremost  step is to generate a NOON state with about hundred photons. We now discuss our proposal for this low-NOON generation via a new type of photon gun.
The low-NOON gun, as we  call it, is based on strong atom-light interactions offered by cavity QED and collective enhancement of the interaction strength. The system required to obtain the low-NOON state of photons is an ensemble of cold alkali atoms---roughly a few hundred of them trapped inside an optical cavity---such that the spatial extent of the cloud is much smaller than the wavelength of the light interacting with it. The schematic of the device and the detailed level structure of the atoms being targeted is shown in Fig.~\ref{Fig:NOONGun}.
The operational steps required are described below. {\em Step 1:} single-step generation of the GHZ state of atoms in two of its internal state via the Hamiltonian given in Eq.~\eqref{Eq:GHZHamiltonian},
\begin{equation}
H_{\rm GHZ} = \hbar \eta \sum_{j,k=1}^{N} \hat{S}_{j}^{+} \hat{S}_{k}^{-} = \hbar \eta\[ \frac{N}{2}\(\frac{N}{2} +1\)-\hat{S}_z^2 +\hat{S}_z\]\,,
\label{Eq:GHZHamiltonian}
\end{equation}
where $\hat{S}_{j}^+ = \ket{{\rm a}_j}\bra{{\rm b}_j}$, $\hat{S}_{k}^- = \ket{{\rm b}_k}\bra{{\rm a}_k}$,  $S_z=\sum_{j} S_{z,j}=\sum_{j}\ket{{\rm a}_j}\bra{{\rm a}_j}-\ket{{\rm b}_j}\bra{{\rm b}_j}$ and $\eta=\Omega_c g \Delta/(\kappa^2 +\Delta^2)$  is the Raman Rabi frequency, signifying the coupling between the two states $\ket{\rm a}$ and $\ket{\rm b}$, achieved through the vacuum mode of a cavity (with decay rate $\kappa$) and a time-varying classical field (See inset 1 in Fig.~\ref{Fig:NOONGun}). The approach is well studied~\cite{AtomicGHZ} and can be used,  with an initial state of all the atoms being in the superposition $(\ket{\rm a}+\ket{\rm b})/\sqrt{2}$, to generate GHZ states in the basis given by $\{\ket{+}\equiv (\ket{\rm a} + \ket{\rm b})/\sqrt{2}, \ket{-}\equiv (\ket{\rm a} - \ket{\rm b})/\sqrt{2} \}$ after the atom-field evolution for time $t$ such that $\eta t=\pi$. The basis rotation can be readily performed by the application of Raman pulse (of area $\pi/2$) coupling the two states to form the GHZ state,  in the familiar basis $\{\ket{\rm a}, \ket{\rm b} \}$, $(\ket{\rm aaa\dots}+\ket{\rm bbb\dots})/\sqrt{2}$. The initial state of the atoms $(\ket{\rm a}+\ket{\rm b})/\sqrt{2}$ could be prepared by two classical Raman fields coupling the levels, much like the Raman process shown in Fig.~\ref{Fig:NOONGun}.
 Direct implementation in Bose-Einstein condensates (BEC) may also be possible by using the scheme of~\cite{Raghavan:2001}, which employs the interparticle interaction for generation of arbitrary Dicke States within a BEC.
{\em Step 2: Entanglement Transfer.} Once a GHZ state of the  atoms is prepared, a coupled STIRAP process (See Fig.~\ref{Fig:NOONGun}) can be achieved by controlling the time-variation of the pump field Rabi frequency $\Omega_P(t)$  to generate an entangled state of cavity photons: $\ket{N_{\circlearrowleft} 0_{\circlearrowright} :: 0_{\circlearrowleft} N_{\circlearrowright} }$ that is entangled in the two counter-rotating polarization modes. This polarization mode entanglement can be readily converted into path entanglement with the help of simple optical elements as shown in detail in Fig.~\ref{Fig:NOONGun}.

The {\em Step 2}, described above, requires  a coupled STIRAP operation for transfer of entanglement from the GHZ state of the atoms to the polarization modes of the photons. This is the most important result of this letter, which offers a device that we call a NOON gun.  This device, however, can not be directly used to generate arbitrarily large number of path-entangled photons. The atom-light interaction needs to be collective requiring that all the atoms see the same strength for both the cavity field and the classical field. Furthermore, the cavity decay and the atomic spontaneous emission increases as the number of atoms increases. Thus, to limit the integrated noise to less than one photon, within the time required for the adiabatic entanglement-transfer process, the total number of atoms needs to be restricted to about 100. The aforementioned error estimates are studied in great detail by Brown {\it et al.} in the context of a Fock state generator~\cite{Brown:2003} and are directly applicable to our scheme, which can be thought of as a superposition of two Fock state generators. The actual requirement of our bootstrapping approach is to have an entangled state of about 32 photons as a control to boost the experimental phase shift of 0.1 radians to $\pi$ radians. This would require only 32 atoms to be trapped in the cavity, making the scheme highly feasible. An experimental scheme of Sauer {\it et al.}~\cite{Sauer:2004}, for cavity QED with optically transported atoms, allows control over the number of atoms interacting with the cavity field for upto 100 atoms and  could be directly used for implementation of our scheme. In the following, we discuss the physics of our NOON gun in sufficient detail.

The initial state of the atomic cloud is the GHZ state with the component states $\ket{\rm a}$ and $\ket{\rm b}$ being the hyperfine sublevels $m_F=-1$ and $m_F=1$ of the $F=1$ hyperfine manifold of an alkali atom respectively. This atomic cloud is then trapped in a cavity such that the cloud size is much smaller than the wavelength of the fundamental mode of the cavity. Also the external pumping field is assumed to couple to all the atoms identically. These restrictions could be easily obtained within the current experimental parameters of the optical cavity, when the number of atoms is confined to about hundred. The interaction of the atomic cloud with the two polarization modes of the cavity and the $\pi$-polarized pump field can be described by the Hamiltonian
\begin{widetext}
\begin{align}
H&=\hbar \sum_i \Omega_P(t) (\ket{\rm  a'}_i \bra{\rm a}_i  +  \ket{\rm b'}_i\bra{\rm b}_i + \ket{\rm  a}_i \bra{\rm a'}_i  +  \ket{\rm b}_i\bra{\rm b'}_i) + g_L (\hat{c}_{L} \ket{\rm a'}_i \bra{\rm g}_i + \hat{c}_L^{\dagger} \ket{\rm g}_i \bra{\rm a'}_i) + g_R ( \hat{c}_R \ket{\rm b'}_i \bra{\rm g}_i + \hat{c}_R^{\dagger}\ket{\rm g}_i \bra{\rm b'}_i ) 
\label{Eq:NOONGunHam1}
 \\
&=\hbar \left[\Omega_P(t) (\hat{d}_{\rm a'}^\dagger \hat{d}_{\rm a} + \hat{d}_{\rm a}^\dagger \hat{d}_{\rm a'} + \hat{d}_{\rm b'}^\dagger \hat{d}_{\rm b} + \hat{d}_{\rm b}^\dagger \hat{d}_{\rm b'}) + g_L(\hat{c}_L \hat{d}_{\rm a'}^\dagger \hat{d}_{\rm g}  + \hat{c}_L^{\dagger} \hat{d}_{\rm g}^\dagger \hat{d}_{\rm a'}) + g_R (\hat{c}_R \hat{d}_{\rm b'}^\dagger \hat{d}_{\rm g}  + \hat{c}_R^{\dagger} \hat{d}_{\rm g}^\dagger \hat{d}_{\rm b'} )\right]\,,
\label{Eq:NOONGunHam2}
\end{align}
\end{widetext}
where $\hat{c}_{L,R}$ and $\hat{c}_{L,R}^{\dagger}$ are the cavity mode photon annililation and creation operators for the left ($L$) and right ($R$) circular modes of polarization and $i$ is the label for the atoms. 
We arrived at the second line (Eq.~\eqref{Eq:NOONGunHam2}) by introducing number representation for the collective atomic states and the corresponding operators $\hat{d}$ and $\hat{d}^\dagger$ labeled by the appropriate atomic level signifying annihilation or creation of atom in state $d$. 

It can be readily realized that this system contains two coupled $\Lambda$ systems. It is important to identify quantities conserved under the action of the Hamiltonian in order to understand the form of the eigenstates of the system. The conserved quantities are the total number of atoms in the five states, $N=\sum_i \hat{d}_{i}^\dagger \hat{d}_{i}$ with $i\in\{\rm a, a', b, b', g\}$, and the quantity $M=\hat{d}_{g}^\dagger \hat{d}_{g} - \hat{c}_{L}^\dagger \hat{c}_{L} - \hat{c}_{R}^\dagger \hat{c}_{R}$.  It should also be noted that the initial state in the given manifold, labeled by $N$ and $M$,  shall always remain in that manifold under the action of the Hamiltonian. Moreover, each manifold contains a dark state, which does not contain any atoms in the excited states $\ket{\rm a'}$ and $\ket{\rm b'}$. We do not give the complete form of the general dark state; however, we note an important observation that is useful for the problem at hand. If the initial state of the cavity fields is chosen such that $\hat{c}_L^{\dagger} \hat{c}_L = \hat{c}_R^{\dagger}\hat{c}_R=0$ and the atomic state is such that all the atoms are in the $\Lambda$-type system formed by either  $\ket{\rm g}-\ket{\rm b'}-\ket{\rm b}$ or $\ket{\rm g}-\ket{\rm a'}-\ket{\rm a}$ (e.g. the GHZ state)  then the interaction Hamiltonian reduces to either $ H^{(1)}_{\rm eff} = \hbar \left[\Omega_P(t) (\hat{d}_{\rm b'}^\dagger \hat{d}_{\rm b} + \hat{d}_{\rm b}^\dagger \hat{d}_{\rm b'})  + g_R (\hat{c}_R \hat{d}_{\rm b'}^\dagger \hat{d}_{\rm g}  + \hat{c}_R^{\dagger} \hat{d}_{\rm g}^\dagger \hat{d}_{\rm b'} )\right] $ or 
$ H^{(2)}_{\rm eff} = \hbar \left[\Omega_P(t) (\hat{d}_{\rm a'}^\dagger \hat{d}_{\rm a} + \hat{d}_{\rm a}^\dagger \hat{d}_{\rm a'}) + g_L(\hat{c}_L \hat{d}_{\rm a'}^\dagger \hat{d}_{\rm g}  + \hat{c}_L^{\dagger} \hat{d}_{\rm g}^\dagger \hat{d}_{\rm a'}) \right] $ as the effective Hamiltonian governing the system dynamics. This gives two different manifolds of states with the conserved quantities $N$ and $M^{(1)}=\hat{d}_{g}^\dagger \hat{d}_{g}  - \hat{c}_{R}^\dagger \hat{c}_{R}$ or $M^{(2)}=\hat{d}_{g}^\dagger \hat{d}_{g} - \hat{c}_{L}^\dagger \hat{c}_{L} $. It is clear that these manifolds to not couple to each other and the corresponding dark states are given by
\begin{align}
\ket{\Psi^{(1)}} &=\frac{1}{D} \sum_{j=0}^{N} \frac{(-\Omega_P(t)/g_L)^j}{\sqrt{(N-j)! j! j!}} \ket{(N-j)_{\rm a}, 0_{\rm b}, j_{\rm g}, j_L, 0_R}  \\
\ket{\Psi^{(2)}} &= \frac{1}{D'} \sum_{k=0}^{N} \frac{(-\Omega_P(t)/g_R)^k}{\sqrt{(N-k)! k! k!}} \ket{0_{\rm a}, (N-k)_{\rm b}, k_{\rm g}, 0_L, k_R}
\end{align}
where $D$ and $D'$ are the appropriate normalization constants.
These specialized dark states of a three-level $\Lambda$-type system
for a collective atomic ensemble simultaneously coupled to a quantized and a classical field are well studied in Refs.~\cite{Lukin:2000,Brown:2003}.
\forget{The cavity is assumed to contain no photons before the evolution starts. In this case the system can be shown to have the following set of degenerate dark state-manifolds of the form $\{\ket{n_{\rm a}-j, 0_b, n_{\rm g}+j, n_L+j, 0_R}, \ket{0_{\rm a}, n_{\rm b}-l, n_{\rm g}+k, 0_L, n_R+k}\}$ where $j\in\{0,n_{\rm a}\} \mbox{ and } k\in\{0,n_{\rm b}\}$ that are decoupled from each other.
Note that the reason behind the decoupling of these two dark state manifolds is the form of the initial state of the atoms and vacuum state of the cavities. Note that the state where there are no atoms in level b(a) and no photons in the $R$($L$) mode of the cavity field do not evolve under the action of the Hamiltonian in Eq.~\eqref{Eq:NOONGunHam2}.
The actual form of the  Dark states discussed in the above manifold are given by
\begin{align}
\ket{\Psi^{(1)}} &=\frac{1}{D} \sum_{j} \(\frac{\Omega_P(t)}{g_L}\)^j \ket{n_{\rm a}-j, 0_b, n_{\rm g}+j, n_L+j, 0_R}  \\
\ket{\Psi^{(2)}} &= \frac{1}{D'} \sum_{k} \(\frac{\Omega_P(t)}{g_R}\)^k \ket{0_{\rm a}, n_{\rm b}-l, n_{\rm g}+k, 0_L, n_R+k}
\end{align}}
The state notation for the complete atom-cavity field states is self-explanatory, where the excited levels $\ket{\rm a'}$ and $\ket{\rm b'}$ are not shown as they are not occupied. It is imperative to point out at this place is that the dark states $\ket{\Psi^{(1)}}$ and $\ket{\Psi^{(2)}}$ are dynamically decoupled from each other in the sense that if the initial state contains a certain proportion of both the states, that proportion shall be left unchanged by the evolution.  The evolution or relative dominance of the components of the dark states can be controlled by controlling the time evolution of the pump field via the Rabi frequency $\Omega_P(t)$. Thus,  the adiabatic transformations $\ket{N_{\rm a},0_{\rm b},0_{\rm g}, 0_L,0_R}\rightarrow\ket{0_{\rm a},0_{\rm b},N_{\rm g}, N_L,0_R}$ and $\ket{0_{\rm a},N_{\rm b}, 0_{\rm g}, 0_L, 0_R}\rightarrow \ket{0_{\rm a},0_{\rm b}, N_{\rm g}, 0_L, N_R}$ can be obtained deterministically.  The result is such that the two components of the initially prepared atom-cavity state, $(\ket{N_a,0_b,0_g, 0_L,0_R} + \ket{0_a,N_b, 0_g, 0_L, 0_R})/\sqrt{2}$ evolve independently into the state 
$(\ket{0_a,0_b,N_g, N_L,0_R} + \ket{0_a,0_b, N_g, 0_L, N_R})/\sqrt{2}$,  just by adiabatic increase of the pump-field intensity such that $\Omega_p(t)\gg g_L,g_R$ in the long time limit like in the STIRAP processes. Moreover, choosing the detuning such that $\Delta \gg \Omega_P(t)$ guarantees that the spontaneous emission noise from the upper levels and spurious absorption events are avoided within the interaction time.
The final state of all the atoms is $\ket{\rm g}$ and the field state can be written in an abbreviated manner as 
$\ket{N_{\circlearrowleft} 0_{\circlearrowright} :: 0_{\circlearrowleft} N_{\circlearrowright} }\equiv
\ket{N_{\circlearrowleft} 0_{\circlearrowright}} + \ket{0_{\circlearrowleft} N_{\circlearrowright} }/\sqrt{2}$. This polarization entangled state of photons can be readily converted into path entangled NOON state in a chosen polarization mode by outcoupling it and passing through simple optical elements as depicted in Fig.~\ref{Fig:NOONGun}. The polarization NOON state itself can be used for  Heisenberg-limited measurement of polarization angle shifts such as is exploited in magnetometry~\cite{Kuzmich:1998}. Physically, the above-mentioned NOON gun is similar in operation to the experimentally demonstrated deterministic single photon source~\cite{Kuhn:2002} and a recent theoretical proposal for Fock-state generation of Ref.~\cite{Brown:2003}. \forget{In fact, the NOON gun proposed here could be thought of as two coupled Fock-State generators,  which when fed with atoms in GHZ state facilitate generation of a NOON state.  The difference comes in the initial state of the atoms being entangled and a superposition of two STIRAP processes which are completely identical to each other but dynamically decoupled from each other provided the system is in certain initial states. Essentially, the device performs complete entanglement transfer from the atoms to the photons leaving all the  atoms in an identical final state.} 

\begin{figure}[ht]
\includegraphics[width=0.77\columnwidth]{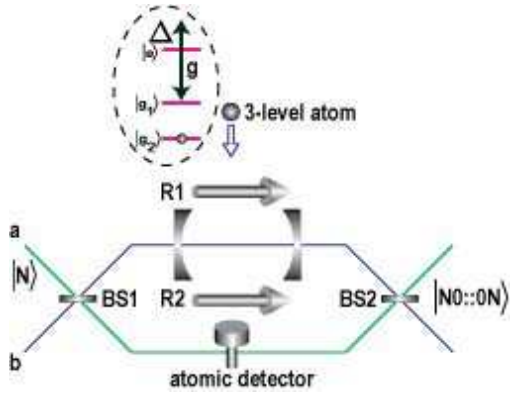}
\caption{\label{Fig:cavityQED-QFG} Quantum Fredkin gate based on nonlinearity obtainable via cavity QED and Ramsey Interferometry}
\end{figure}

This completes the discussion of the {\em bootstrapping} procedure for high-NOON generation. Now we briefly discuss a different implementation of the Kerr nonlinearity based on the atom-cavity dispersive interaction~\cite{TheBook} that can be used in place of the  cross-Kerr nonlinearities obtained via quantum coherence effects. The scheme is depicted in Fig.~\ref{Fig:cavityQED-QFG} where a cavity is introduced in the path of the optical mode b after the beam splitter BS1. In comparison with the scheme of Fig.~\ref{Fig:QFG}(a), the new scheme (See Fig.~\ref{Fig:cavityQED-QFG}) uses a Ramsey Interferometer in place of the MZI-2.
The mathematical transformation of the Quantum Fredkin Gate is given by~\cite{Gerry:2001} as
$
\hat{U}_{\rm QFG} = \exp(\ri \chi \hat{c}^\dagger \hat{c} \hat{J}_0) \exp(\ri  \chi \hat{c}^\dagger \hat{c} \hat{J}_2)\,,
$
where $\hat{J}_0=(\hat{a}^\dagger \hat{a} + \hat{b}^\dagger \hat{b})/2$ and $\hat{J}_2=(\hat{a}^\dagger \hat{b} - \hat{a} \hat{b}^\dagger)/2\ri$ are the Schwinger angular momentum representations of the photonic operators and $\chi=\pi$ is the cross-Kerr nonlinearity required for generation of the NOON state at the output of  MZI-1 after detection of the control photon in the one of the detectors at the output ports of MZI-2 [See Fig.~\ref{Fig:QFG}(a)].
The Ramsey interferometer that replaces MZI-2 consists of three regions: R1 and R2 are Ramsey classical field zones giving transformations $\ket{{\rm g}_1} \rightarrow (\ket{{\rm g}_1}+\ket{{\rm g}_2})/\sqrt{2} \mbox{ and } \ket{{\rm g}_2} \rightarrow (\ket{{\rm g}_1} - \ket{{\rm g}_2})/\sqrt{2}$ with a strong cavity dispersively coupled (large detuning $\Delta$)  with the atomic transition $\ket{{\rm g}_1}$ to $\ket{{\rm e}}$ in between the two Ramsey zones. The dispersive interaction with the cavity imparts a phase shift of $g^2 \tau_c/\Delta$ to the atomic level $\ket{{\rm g}_1}$ and no phase shift to state $\ket{{\rm g}_2}$. Thus, after the passage of an atom initially prepared in state $\ket{{\rm g}_2}$ when the photons in mode ${\rm b}$ are confined in the cavity we obtain the following transformation
$
\hat{U}_{\rm Ramsey-QFG} = \exp[\ri (g^2 \tau_c/\Delta) \hat{\sigma}_+ \hat{\sigma}_- \hat{J}_0] \exp[\ri (g^2 \tau_c/\Delta) \hat{\sigma}_+ \hat{\sigma}_- \hat{J}_2]\,.
$ 
Here $\hat{\sigma}_+=\ket{{\rm g}_1}\bra{{\rm g}_2}$ and $\hat{\sigma}_-=\ket{{\rm g}_2}\bra{{\rm g}_1}$ are the atomic operators, $g$ is the atom-cavity interaction strength and $\tau_c$ is the time atom spends in the cavity. With $g^2 \tau_c/\Delta=\pi$, the evolution of the system is identical to the NOON generation scheme of Ref.~\cite{Gerry:2001} such that detection of the atom,  as it comes out of the cavity, in the state $\ket{{\rm g}_1}$ or $\ket{{\rm g}_2}$, gives out the NOON state in form
$
\ket{\Psi_{1(2)}}_{\rm ab} = \left[ \ket{N}_{\rm a} \ket{0}_{\rm b} \pm \re^{-\ri N \pi/2} \ket{0}_{\rm a} \ket{N}_{\rm b}\right]/2
$
at the output of the Mach-Zehnder interferometer. The nonlinearity in this setup is completely controllable via the atomic velocity, that is, the atomic passage time through the cavity $\tau_c$, and the cavity parameters $g$ and $\Delta$. This makes it straightforward to obtain the phase shift of the order of $\pi$ with just one photon present in the cavity. By introducing a delay in the path of mode a the time spent by photons in mode b inside the cavity can be trivially compensated to obtain a balanced MZI. Within the experimental optical cavity QED parameters $g \tau_c$ can be as large as $10^5$~\cite{Kuhn:2002}. With $\Delta=10 g$ and $g^2 \tau_c/\Delta=\pi$ we need a moderate value $g \tau_c = 10 \pi$, which is much easier to obtain as it requires less cooling of the atoms entering the cavity. For  $g=5\pi$ MHz, the atom-cavity interaction time is $\tau_c=2 \mu$s, which is much smaller than the photon lifetime (few tens of miliseconds) in the cavity; thus, the approach presented here is well within the current experimental parameters.

To summarize, we have devised a {\em bootstrapping} approach to NOON-state generation for photons based on Quantum Fredkin gate via atom coherence effect based cross-Kerr nonlinearities. In the process we have proposed a device that can produce NOON states of up to one hundred photons on demand in a deterministic manner. Furthermore, we have devised a scheme for NOON-state generation based on Ramsey interferometry and a cavity-enhanced Kerr nonlinearity.  It is assumed that the Fock states of arbitrarily large number of photons are available as inputs; which could be generated by applying  the proposal of Ref.~\cite{Brown:2003} when applied to cold Rydberg atoms trapped in microwave cavities or via a QND photon number measurement of a coherent state containing large number of photons on average. Our strong hope is that the ideas presented here shall simulate a growth of experimental activity in generation of entangled photon states with larger and larger number of photons. 

 We would like to acknowledge support from the Hearne institute, the Disruptive Technologies Office and the Army Research Office.


\begin{thebibliography}{10}

\bibitem{QL}
A.~N. Boto {\it et~al.}, Phys. Rev. Lett. {\bf 85},  2733  (2000);
P. Kok {\it et~al.}, Phys. Rev. A {\bf 63},  063407  (2001).

\bibitem{HLI}
M.~J. Holland and K. Burnett, Phys. Rev. Lett. {\bf 71},  1355  (1993); J.~J. Bollinger {\it et~al.}, Phys. Rev. A {\bf 54},  R4649  (1996); 
J.~P. Dowling, Phys. Rev. A {\bf 57},  4736  (1998);
Z.~Y. Ou, Phys. Rev. A {\bf 55},  2598  (1997); 
R.~A. Campos, C.~C. Gerry, and A. Benmoussa, Phys. Rev. A {\bf 68},  023810
  (2003).

\bibitem{NOONExp}
M.~W. Mitchell, J.~S. Lundeen, and A.~M. Steinberg, Nature (London) {\bf 429},
  161  (2004);
P. Walther {\it et~al.}, {\it ibid.},  158  (2004).

\bibitem{Resch:2005}
K.~J. Resch {\it et~al.}, Phys. Rev. Lett. {\bf 98}, 223601 (2007).

\bibitem{Kok:2002}
P. Kok, H. Lee and J.~P. Dowling, Phys. Rev. A {\bf 65},  052104  (2002).

\bibitem{EITPhysicsToday}
S.~E. Harris, Physics Today {\bf 50},  36-42 (1997).

\forget{\bibitem{Bootstrapping} {\em Bootstrapping}, in the context of computing, means building complex tools after creation of simple tools that allow for creation of more complex tools. This suggests a two-step approach we are taking for arbitrary NOON state generation.}

\bibitem{Gerry:2001}
C.~C. Gerry and R.~A. Campos, Phys. Rev. A {\bf 64},  063814  (2001).

\bibitem{PhaseoniumCrossKerr}
M.~D. Lukin and A. Imamoglu, Phys. Rev. Lett. {\bf 84} 1419 (2000);
D.~Petrosyn and G. Kurizki, Phys. Rev. A {bf 65} 033833 (2002).

\bibitem{CrossKerrExp}
H. Kang and Y. Zhu, Phys. Rev. Lett. {\bf 91},  093601  (2003);
%
Z.-B. Wang, K.-P. Marzlin, and B.~C. Sanders, Phys. Rev. Lett. {\bf 97}, 063901  (2006).
  
\bibitem{PhotonNumberResolvingDetectors}
A. Imamoglu,  Phys. Rev. Lett. {\bf 89}, 163602 (2002); D.~F.~V. James, P.~G. Kwiat, Phys. Rev. Lett. {\bf 89}, 183601 (2002); W.~J. Munro, et al., Phys. Rev. A  {\bf 71}, 033819 (2005); D. Rosenberg, et al., Phys. Rev. A 71, 061803(R) (2005); K.~T. Kapale, J. Mod. Opt. {\bf 54}, 327 (2007).

\bibitem{AtomicGHZ}
G.~S. Agarwal, R.~R. Puri, and R.~P. Singh, Phys. Rev. A {\bf 56},  2249 (1997);
S.-B. Zheng, Phys. Rev. Lett. {\bf 87},  230404  (2001).

\bibitem{Raghavan:2001}
S. Raghavan {\it et~al.}, Opt. Commun. {\bf 188},  149
   (2001).

\bibitem{Brown:2003}
K.~R. Brown {\it et~al.}, Phys. Rev. A {\bf 67},  043818  (2003).

\bibitem{Sauer:2004}
J.~A. Sauer {\it et al.}, Phys. Rev. A {\bf 69}, 051804(R) (2004).

\bibitem{Lukin:2000}
M.~D. Lukin, S.~F. Yelin, and M. Fleischhauer, Phys. Rev. Lett. {\bf 84},  4232 (2000).

\bibitem{Kuzmich:1998} 
A.~Kuzmich and L.~Mandel, Quantum and Semiclass. Opt. {\bf 10}, 493 (1998).

\bibitem{Kuhn:2002}
A. Kuhn, M. Hennrich, and G. Rempe, Phys. Rev. Lett. {\bf 89},  067901  (2002).

\bibitem{TheBook}
M.~O. Scully and M.~S. Zubairy {\it Quantum Optics}, pg. 547--554 Cambridge University Press, Oxford (1997).
\end{thebibliography}

\end{document}